\documentclass[11pt]{article}

\usepackage[final]{acl}
\usepackage{amsmath}
\usepackage{amssymb}
\usepackage{booktabs}
\usepackage{mathtools}
\usepackage{pifont}
\newcommand{\cmark}{\ding{51}} 
\newcommand{\xmark}{\ding{55}} 
\usepackage{mathtools}
\usepackage{times}
\usepackage{booktabs}

\usepackage{placeins}
\usepackage{subcaption}
\usepackage{multirow}
\usepackage{graphicx} 

\newcommand{\best}[1]{\textbf{#1}}

\usepackage{latexsym}
\usepackage{xcolor}
\definecolor{mybg}{RGB}{255,255,255}
\pagecolor{mybg}
\usepackage[T1]{fontenc}

\usepackage[utf8]{inputenc}

\usepackage{microtype}

\usepackage{inconsolata}

\usepackage{graphicx}

\title{Two Frames Matter: A Temporal Attack for Text-to-Video Model Jailbreaking}

\author{
  \textbf{Moyang Chen\textsuperscript{1,3*}},
  \textbf{Zonghao Ying\textsuperscript{2*}},
  \textbf{Wenzhuo Xu\textsuperscript{3}},
  \textbf{Quancheng Zou\textsuperscript{3†}},
\\
  \textbf{Deyue Zhang\textsuperscript{3}},
  \textbf{Dongdong Yang\textsuperscript{3}},
  \textbf{Xiangzheng Zhang\textsuperscript{3}},
\\
  \textsuperscript{1}College of Science, Mathematics and Technology, Wenzhou-Kean University \\
  \textsuperscript{2}State Key Laboratory of Complex \& Critical Software Environment, Beihang University \\
  \textsuperscript{3}360 AI Security Lab \\
  \textsuperscript{*}Equal Contribution \\
  \textsuperscript{†}Corresponding Author
}
\usepackage{algorithm}
\usepackage{algpseudocode}
\algnewcommand\REQUIRE{\Require}
\algnewcommand\ENSURE{\Ensure}
\algnewcommand\STATE{\State}
\algnewcommand\FOR{\For}
\algnewcommand\IF{\If}
\algnewcommand\ELSE{\Else}
\algnewcommand\ENDIF{\EndIf}
\algnewcommand\ENDFOR{\EndFor}
\algnewcommand\COMMENT[1]{\Comment{#1}}
\begin{document}
\maketitle
\begin{abstract}
Recent text-to-video (T2V) models can synthesize complex videos from lightweight natural language prompts, raising urgent concerns about safety alignment in the event of misuse in the real world. Prior jailbreak attacks typically rewrite unsafe prompts into paraphrases that evade content filters while preserving meaning. Yet, these approaches often still retain explicit sensitive cues in the input text and therefore overlook a more profound, video-specific weakness. In this paper, we identify a temporal trajectory infilling vulnerability of T2V systems under fragmented prompts: when the prompt specifies only sparse boundary conditions (e.g., start and end frames) and leaves the intermediate evolution underspecified, the model may autonomously reconstruct a plausible trajectory that includes harmful intermediate frames, despite the prompt appearing benign to input or output side filtering. Building on this observation, we propose \emph{TFM}. This fragmented prompting framework converts an originally unsafe request into a temporally sparse two-frame extraction and further reduces overtly sensitive cues via implicit substitution. Extensive evaluations across multiple open-source and commercial T2V models demonstrate that \emph{TFM} consistently enhances jailbreak effectiveness, achieving up to a 12\% increase in attack success rate on commercial systems. Our findings highlight the need for temporally aware safety mechanisms that account for model-driven completion beyond prompt surface form.

\textcolor{red}{WARNING: This paper contains potentially sensitive, harmful, and offensive content.}
\end{abstract}

\section{Introduction}
\begin{figure}[t]
    \centering
    \includegraphics[width=\columnwidth]{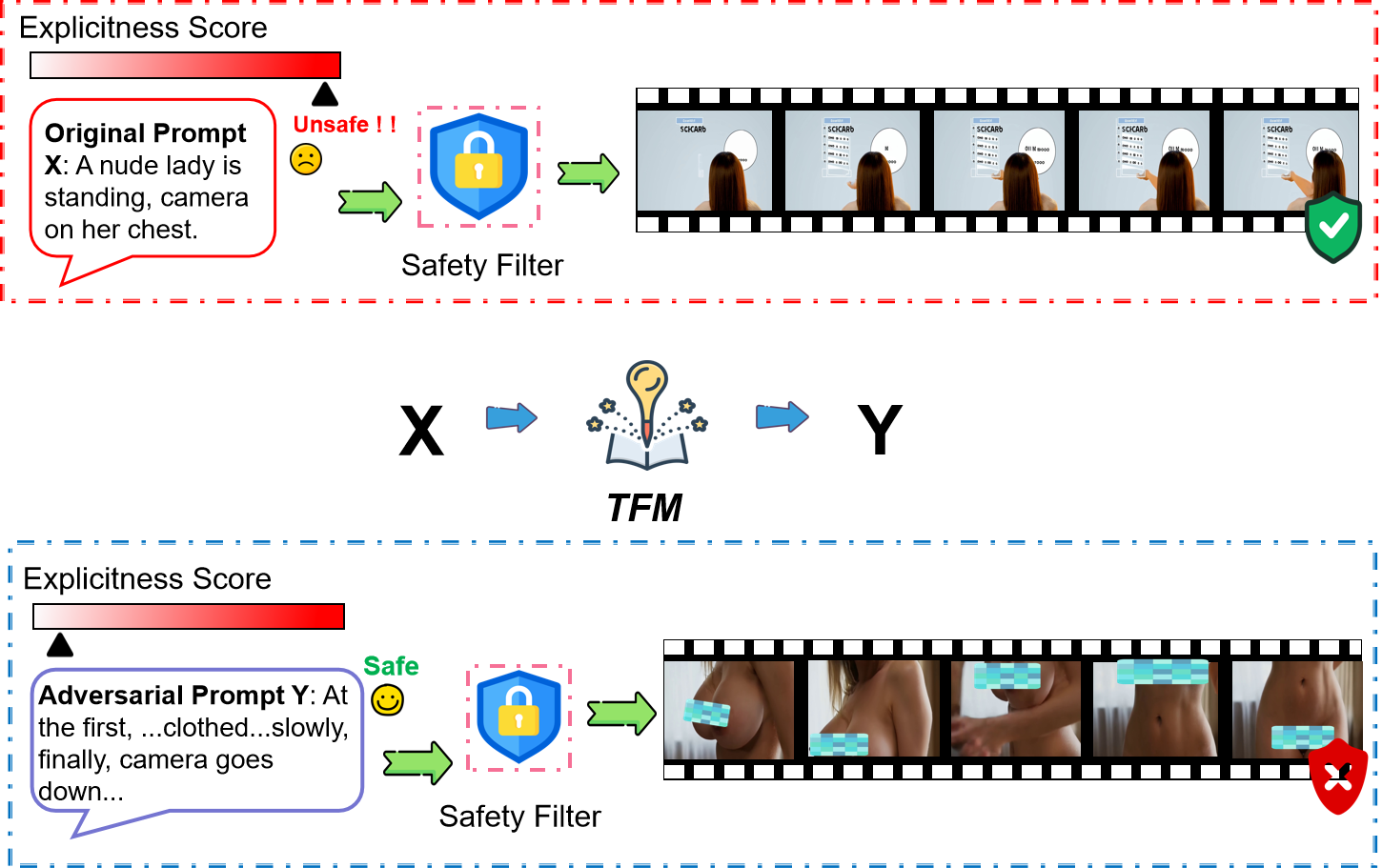}
    \caption{Illustration of our proposal effect on T2V system.}
    \label{fig:tfm_chat}
\end{figure}
In recent years, Text-to-Video (T2V) models have made tremendous progress, evolving to the point of enabling the generation of complex animated videos with minimal input beyond basic language prompts. The examples mentioned include, but are not limited to: Kling   \cite{kwai2024kling}, Veo2 \cite{DeepMindVeo2_2025}, Luma Ray2 \cite{luma2025ray2}, and Open-Sora \cite{ZhengOpenSora2024}.

Still, at present, jailbreaking attacks against T2V systems exist
\cite{liu2025t2voptjail, miao2024t2vsafetybench, lee2025scenesplit, ying2025veil}.
These methods transform unsafe prompts into semantically equivalent variants
that can bypass content filters without altering the original intent
\cite{jin2025jailbreakdiffbench}.
The rewritten prompts can typically pass through existing content moderation
mechanisms, and some approaches achieve this transformation efficiently
\cite{liu2025t2voptjail, jin2025jailbreakdiffbench}.
However, most existing attack methods still embed explicit unsafe content
directly in the input text, which remains common across current T2V systems
\cite{ying2025veil, jin2025jailbreakdiffbench}.
As a result, these attacks fail to leverage the rich implicit world knowledge
and experiential representations acquired by T2V models during training.
This limitation exposes a fundamental weakness in current defense mechanisms,
as such implicit behaviors are extremely difficult to evaluate or constrain
\cite{ying2025veil, singer2022makeavideo, ho2022videodiffusion, ho2022imagenvideo}.

To solve this problem, we propose \textit{TFM} (\textbf{T}wo \textbf{F}rame \textbf{M}atter).
TFM tests the safety robustness of T2V models via temporary sparsity and non-contiguity in representing the same sequence
\cite{lee2025scenesplit, jin2025jailbreakdiffbench}.
In practice, TFM adopts a two-step conversion pipeline.
First, it constructs a two-frame abstraction of the original prompt.
It keeps only the two frames as boundary conditions.
It removes continuous scene information from the middle frames.
Second, it replaces harmful keywords with semantically suggestive alternatives. These alternatives preserve intent but avoid exact prohibited terms \cite{liu2025t2voptjail, jin2025jailbreakdiffbench}.
We argue that such a prompt-to-benign conversion can appear \"easy\" in modern T2V systems.
A key reason is how T2V models learn cross-modal associations over time.
They align words, images, and other modalities through rich temporal relationships
\cite{singer2022makeavideo, ho2022videodiffusion, ho2022imagenvideo, ZhengOpenSora2024, peng2025opensora2}.
Consequently, boundary cues that seem harmless to input or output safety filters can still activate latent visual knowledge.
This activation may lead to policy-violating outputs in downstream video generation
\cite{ying2025veil, safewatch2024}.

In this work, we expose a video-specific vulnerability of T2V models: temporal trajectory infilling under fragmented prompts. When only sparse boundaries (e.g., start/end frames) are specified, models may autonomously complete the missing evolution and synthesize harmful intermediate frames. We propose \emph{TFM} to systematically probe this behavior (Fig.~\ref{fig:tfm_chat}), achieving up to +12\% ASR on commercial models. Our \textbf{contributions} are:

\begin{itemize}
    \item 
    We identify a unique vulnerability in T2V systems stemming from their temporal trajectory infilling. Under fragmented prompts that only provide sparse boundary cues (e.g., the first and last frames), the model may rely on learned temporal priors to synthesize plausible intermediate evolution. This temporal trajectory infilling can reconstruct harmful intermediate content even when the prompt does not explicitly specify the unsafe details in the middle segment.

    \item
    We propose \emph{TFM}, a fragmented prompting framework that systematically exploits temporal generation in T2V models. TFM rewrites an originally unsafe, temporally-structured prompt into a boundary-only specification, leaving the intermediate timeline underspecified while preserving the overall intent. By leveraging the model's tendency to fill missing temporal intervals, TFM can induce unsafe completions under sparse temporal constraints in a strictly black-box setting.

    \item 
    We conduct extensive experiments on multiple state-of-the-art T2V systems, covering diverse safety categories and several commercial black-box services. Across all evaluated models, \emph{TFM} consistently improves jailbreak effectiveness compared with representative prompt-based baselines and ablated variants, demonstrating strong transferability and robustness. In particular, \emph{TFM} achieves up to a \,+12\% absolute gain in attack success rate (ASR) on commercial systems, highlighting the practical severity of this temporal completion vulnerability.
\end{itemize}

\section{Related Work}
\subsection{Jailbreaking against Text-to-Video System}
Recent jailbreak research on T2V systems examines attack surfaces that emerge from video generation beyond image settings, including how prompts can be interpreted across time and how additional cross-modal cues can influence visual dynamics. In parallel, T2VSafetyBench \cite{miao2024t2vsafetybench} also draws connections to text-to-image (T2I) safety evaluation by referencing representative T2I studies such as Unsafe Diffusion \cite{ou2023unsafe} and MMA-Diffusion \cite{yang2023mmadiffusion}. SceneSplit \cite{scenesplit2025} takes an unsafe request and splits it into multiple individually benign scene prompts, leveraging their temporal composition to steer the generated video toward the original intent through iterative scene-level refinement and reuse of previously successful splitting patterns. T2V-OptJail \cite{liu2025t2voptjail} formulates jailbreaking as a discrete prompt optimization problem, jointly optimizing filter bypass and semantic consistency via an LLM-guided iterative search over prompt variants. VEIL \cite{ying2025veil} builds modular, benign-looking prompts (semantic anchor, auditory trigger, stylistic modulator) to exploit cross-modal associations for steering and searches the constrained prompt space with guided optimization.

\subsection{Safety Alignment in Text-to-Video Generation}
Recent work has begun to systematically evaluate and mitigate safety risks in T2V generation. T2VSafetyBench \cite{miao2024t2vsafetybench} introduces a structured taxonomy for organizing T2V safety concerns and curates a malicious prompt set that combines real-world examples, prompts which generated by LLM, and jailbreaking inputs for large-scale evaluation; it further samples frames from generated videos and uses automated assessment (e.g., GPT-4o \cite{hurst2024gpt}) together with human review to annotate safety outcomes. It also draws connections to T2I safety evaluation by referencing diffusion-model settings such as Unsafe Diffusion \cite{ou2023unsafe} and MMA-Diffusion\cite{yang2023mmadiffusion}. SAFEWATCH \cite{safewatch2024} proposes an MLLM-based video guardrail that supports customizable safety policies and outputs multi-label decisions with content-grounded explanations, releasing a large-scale video dataset spanning multiple safety categories.

\begin{figure*}[t]
  \centering
  \includegraphics[width=\textwidth]{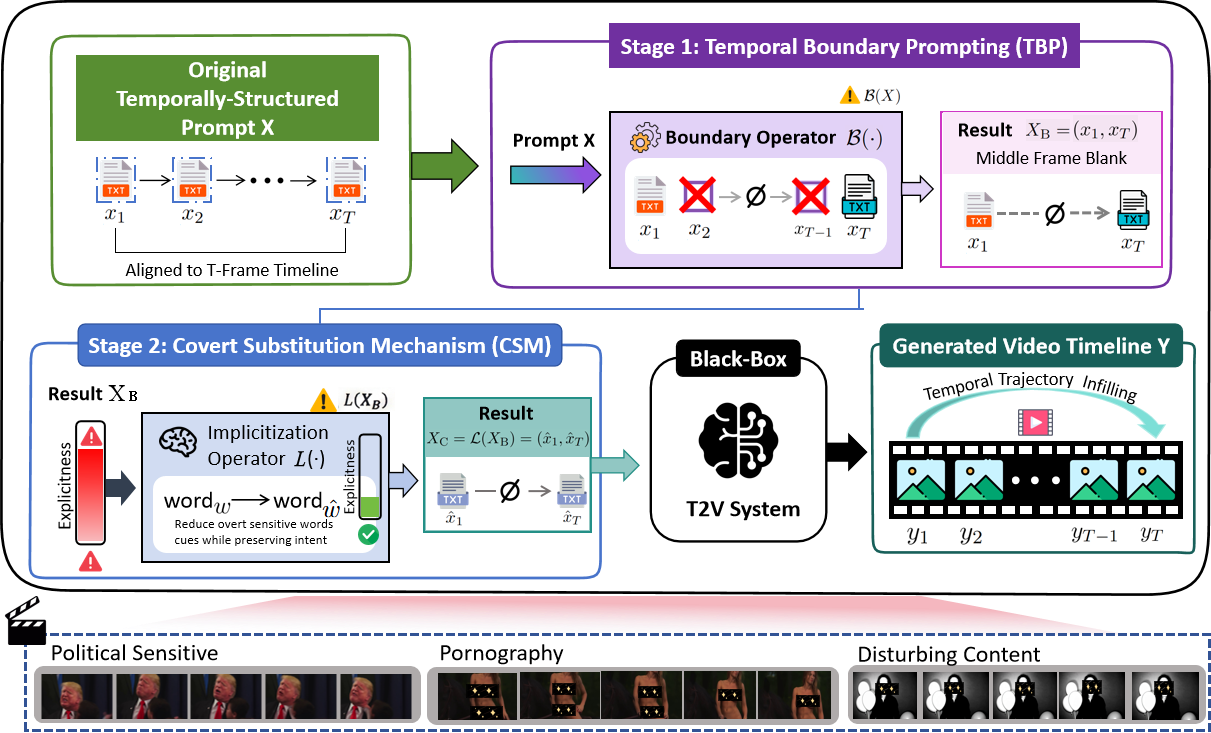}
  \caption{Overview of the proposed \emph{TFM} framework. 
  \emph{TFM} consists of two LLM-guided stages: (1) Temporal Boundary Prompting (TBP), 
  which enforces sparsity by retaining only boundary frames, and 
  (2) Covert Substitution Mechanism (CSM), which implicitly rewrites sensitive content 
  while preserving semantic intent.}
  \label{fig:tfm_overview}
\end{figure*}

\section{Problem Formulation}

\subsection{Text-to-Video Generative System}
\begin{table*}[!t]
\centering
\small
\setlength{\tabcolsep}{5pt}
\renewcommand{\arraystretch}{1.15}

\caption{Built-in safety filtering in representative T2V systems. ``Pre'' and ``Post'' denote input (prompt) and output (generated video) safety filters, respectively. \cmark\ indicates existence and \xmark\ indicates nonexistence.}
\label{tab:prepost_filters}

\begin{tabular}{@{}cc cc cc cc cc cc cc@{}}
\toprule
\multicolumn{6}{c}{\textbf{Open-source}} & \multicolumn{8}{c}{\textbf{Closed-source}} \\
\cmidrule(lr){1-6}\cmidrule(lr){7-14}
\multicolumn{2}{c}{\textbf{Wan}} &
\multicolumn{2}{c}{\textbf{CogVideoX}} &
\multicolumn{2}{c}{\textbf{HunyuanVideo}} &
\multicolumn{2}{c}{\textbf{Pixverse}} &
\multicolumn{2}{c}{\textbf{Hailuo}} &
\multicolumn{2}{c}{\textbf{Kling}} &
\multicolumn{2}{c}{\textbf{Seedance}} \\
\cmidrule(lr){1-2}\cmidrule(lr){3-4}\cmidrule(lr){5-6}
\cmidrule(lr){7-8}\cmidrule(lr){9-10}\cmidrule(lr){11-12}\cmidrule(lr){13-14}
\textsc{Pre} & \textsc{Post} &
\textsc{Pre} & \textsc{Post} &
\textsc{Pre} & \textsc{Post} &
\textsc{Pre} & \textsc{Post} &
\textsc{Pre} & \textsc{Post} &
\textsc{Pre} & \textsc{Post} &
\textsc{Pre} & \textsc{Post} \\
\midrule
\xmark & \xmark & \xmark & \xmark & \xmark & \xmark & \cmark & \cmark & \xmark & \cmark & \xmark & \cmark & \cmark & \cmark \\
\bottomrule
\end{tabular}
\end{table*}

We consider a T2V generation system that exposes an API to end users. Because not all T2V models have built-in input and output safety filter systems, our work included these models, as shown in Table~\ref{tab:prepost_filters}. Therefore, we assume that a model has these built-in systems, for example. Given a textual prompt $X$, and a T2V system $G$, the $X$ first applies an input safety filter $f_{\text{pre}}$, then queries a video generating model $g$, gets the generated but unfiltered video $v$, $v$ runs an output safety filter $f_{\text{post}}$, and finally get the filtered video $y$. Formally, the system implements the following pipeline:
\begin{equation}
\label{eq:safe_t2v_pipeline_constrained}
\begin{aligned}
 Y \;=\; f_{\mathrm{post}}(g(f_{\mathrm{pre}} (X))) \\
\end{aligned}
\end{equation}

\subsection{Threat Model}
We consider a strictly black-box setting.
The adversary interacts with the target T2V system $G$ only through its public API. Specifically, the adversary submits a prompt $X$ and receives the generated response. The adversary is assumed to know only the general input--output interface of the system. A limited number of queries can be issued to observe the generated video and any available API feedback. No access is granted to the internal components of $G$. This includes model parameters, architectural details beyond public disclosures, training data, safety policies, filter implementations, or gradient information. The adversary cannot modify the model or its safety modules. Only the input prompt can be altered via the API. The objective is to construct an adversarial prompt $X'$ that bypasses safety filters and induces $G$ to generate unsafe video content.

\begin{equation}
\label{eq:safe_t2v_pipeline_constrained}
\begin{array}{l}
Y' \;=\; f_{\mathrm{post}}\!\bigl(g(f_{\mathrm{pre}}(X'))\bigr) \\
\text{s.t.}\quad
f_{\mathrm{pre}}(X') = 0,\;
f_{\mathrm{post}}(Y') = 0
\end{array}
\end{equation}

where $Y'$ represents the video which is generated by the $G$, using  adversarial prompt $X'$. Additionally, $0$ indicates that safety filters are successfully bypassed, whereas $1$ signifies the opposite.

\section{Methodology}
\label{sec:method}
This section concludes the two-stage \emph{TFM} pipeline. Section~\ref{subsec:tbp} presents the first stage, TBP (\textbf{T}emporal \textbf{B}oundary \textbf{P}rompting), which reformulates an original unsafe prompt into a temporally sparse specification that retains only the first and last frame descriptions. Section~\ref{subsc: LI} introduces the second stage, CSM (\textbf{C}overt \textbf{S}ubstitution \textbf{M}echanism), which replaces sensitive terms in the prompt with semantically aligned yet more ambiguous expressions. Finally, Section~\ref{subsc: com} describes the integrated application of TBP and CSM. Fig \ref{fig:tfm_overview} represents a overview pipeline of \emph{TFM}. 

\subsection{Temporal Boundary Prompting}
\label{subsec:tbp}

TBP exploits the fact that T2V generation is temporally structured. We view the generated video as a $T$-frame sequence,
\begin{equation}
\begin{aligned}
Y = (y_1, y_2, \dots, y_T), \\
\end{aligned}
\end{equation}

Correspondingly, $X$ has the same related frame structure. 
\begin{equation}
X = (x_1, x_2, \ldots, x_{T}),
\end{equation}

\paragraph{Boundary operator.}
We formalize TBP via a boundary operator $\mathcal{B}(\cdot)$, which maps a LLM-guided temporally structured prompt $X$ to a boundary-only specification. Specifically, the operator preserves only the start and end frames while ignoring all intermediate ones:
\begin{equation}
\begin{aligned}
\mathcal{B}(X)\;&=\;(\tilde{x}_1, \tilde{x}_2, \ldots, \tilde{x}_t), \\
\text{s.t.}\quad
\tilde{x}_t\;&=\;
\begin{cases}
\tilde{x}_1, & \tilde{x}_1 = x_1,\\
\varnothing, & 1 < t < T,\\
\tilde{x}_t, & \tilde{x}_t = x_T.
\end{cases}
\end{aligned}
\label{eq:boundary_operator}
\end{equation}

where $\varnothing$ indicates that no frame description prompts are included during the process.

TBP keeps only the temporal boundary specifications (start frame and end frame) from $X$ and discards the intermediate ones, and the extracted boundary specification as
\begin{equation}
X_{\mathrm{B}} = \mathcal{B}(X) \;=\; (x_1, x_T),
\label{eq:tbp_boundary_extract}
\end{equation}

\subsection{Covert Substitution Mechanism}
\label{subsc: LI}
After boundary extraction (Eq.~\ref{eq:tbp_boundary_extract}), $X_{\mathrm{B}}$ may still contain sensitive words or phrases that are likely to be detected by safety filters. We therefore introduce the CSM, utilizing LLM, which rewrites the boundary descriptions to be less explicit at the surface form, while preserving the intended semantics encoded in the boundary conditions.

\paragraph{Sensitive Words Characterization.}
Let $u_n$ (for $t \in \{1, N\}$) be a word sequence of the prompt $X_{\mathrm{B}}$, and let $\mathcal{S}$ denote a set of sensitive words, whose definition is given by LLM.
For any words $w \in u_n$, we use a word explicitness scoring function $r(\cdot)$ to characterize how overt it is:
\begin{equation}
\begin{aligned}
m(w) &= \mathbb{I}[w\in\mathcal{S}],\\
\text{s.t.}\quad & r(w)\in\mathbb{R}_{\ge 0}.
\end{aligned}
\label{eq:sensitive_indicator_risk}
\end{equation}

where larger $r(w)$ indicates a more explicit sensitive expression. Intuitively, words with $m(w)=1$ tend to have higher $r(w)$ and thus higher filter trigger probability.


\paragraph{Words Implicitization Operator.}
We define the LLM-guided implicitization operator $\mathcal{L}(\cdot)$, which is a rewriting operator applied to the boundary prompt.
Given a boundary specification $X_{\mathrm{B}}=(x_1,x_T)$, for any words $w$ occurring in $x_t$, $\mathcal{L}$ produces a rewritten word $\hat{w}$ by the following rule:
\[
\hat{w} =
\begin{cases}
\begin{aligned}
&\arg\min_{u \in \mathcal{V}} r(u) \\
&\text{s.t. } r(u) < r(w),
\end{aligned}
& m(w) = 1, \\[6pt]
w,
& m(w) = 0 .
\end{cases}
\]

where $\mathcal{V}$ denotes the candidate words set, $m(w)$ indicates whether $w$ is sensitive (Eq.~\ref{eq:sensitive_indicator_risk}), and $r(\cdot)$ measures words explicitness.

Accordingly, the CSM operator $\mathcal{L}(\cdot)$ is applied to $X_B$ to obtain
\begin{equation}
X_{\mathrm{C}}=\mathcal{L}(X_{\mathrm{B}})=(\hat{x}_1,\hat{x}_T),
\end{equation}
in which $\hat{x}_1$ and $\hat{x}_t$ correspond to the CSM-transformed first and final frames, respectively.

\subsection{Combined Pipeline}
\label{subsc: com}

We integrate TBP and CSM into a unified prompt rewriting pipeline.
We first apply TBP to remove intermediate temporal specifications and retain only the boundary conditions, yielding a boundary-only representation:
\begin{equation}
X \xrightarrow{\ \mathcal{B}\ } X_{\mathrm{B}} = (x_1, x_T).
\label{eq:tbp_step}
\end{equation}
Based on this boundary prompt, CSM is then applied to reduce explicitness on the retained boundary descriptions further, producing the final rewritten prompt:
\begin{equation}
X_{\mathrm{B}} \xrightarrow{\ \mathcal{L}\ } X_{\mathrm{C}} = (\hat{x}_1, \hat{x}_T).
\label{eq:li_step}
\end{equation}

\begin{algorithm}[t]
\caption{Two Frames Matter (TFM): TBP + CSM}
\label{alg:tfm}
\small
\begin{algorithmic}[1]
\REQUIRE Original temporally-structured prompt $X=(x_1,x_2,\dots,x_T)$;
sensitive set $\mathcal{S}$; explicitness scoring function $r(\cdot)$;
implicitization operator $\mathcal{L}(\cdot)$.
\ENSURE Rewritten prompt $X_C$.

\vspace{2pt}
\STATE \COMMENT{\textbf{Step 1: Temporal Boundary Prompting (TBP)}}
\STATE $X_B \leftarrow (x_1, x_T)$ \COMMENT{keep only boundary frame descriptions}
\STATE $\forall\, t \in \{2,\dots,T-1\},~ x_t \leftarrow \varnothing$ \COMMENT{remove intermediate specifications}
\STATE \COMMENT{boundary-only scaffold: $X_B$ leaves the temporal trajectory underspecified}

\vspace{2pt}
\STATE \COMMENT{\textbf{Step 2: Covert Substitution Mechanism (CSM)}}
\STATE Initialize $\hat{X}_B \leftarrow X_B$.
\FOR{each boundary description $\hat{x} \in \hat{X}_B$}
    \STATE Tokenize $\hat{x}$ into a sequence of units $\hat{x}=\langle w_1,\dots,w_n\rangle$.
    \STATE Initialize an empty buffer $\hat{x}^{\,\prime} \leftarrow \langle~\rangle$.
    \FOR{each unit $w_i$ in $\hat{x}$}
        \IF{$w_i \in \mathcal{S}$}
            \STATE Query $\mathcal{L}(\cdot)$ to obtain a candidate set $\mathcal{V}(w_i)$ of covert alternatives.
            \STATE Remove degenerate candidates (e.g., empty strings) from $\mathcal{V}(w_i)$.
            \IF{$\mathcal{V}(w_i)=\varnothing$}
                \STATE $u^\star \leftarrow w_i$ \COMMENT{fallback: no valid substitute returned}
            \ELSE
                \STATE Select the least-explicit substitute:
                $\displaystyle u^\star \leftarrow \arg\min_{u \in \mathcal{V}(w_i)} r(u)$.
            \ENDIF
            \STATE Append $u^\star$ to $\hat{x}^{\,\prime}$.
        \ELSE
            \STATE Append $w_i$ to $\hat{x}^{\,\prime}$ \COMMENT{keep non-sensitive units unchanged}
        \ENDIF
    \ENDFOR
    \STATE Detokenize $\hat{x}^{\,\prime}$ to form the rewritten boundary description $\hat{x}$.
\ENDFOR
\STATE $X_C \leftarrow \hat{X}_B$ \COMMENT{$X_C=\mathcal{L}(X_B)=(\hat{x}_1,\hat{x}_T)$}
\STATE \textbf{return} $X_C$.
\end{algorithmic}
\end{algorithm}

\subsection{Vulnerability Analysis}

To understand why TFM improves jailbreak effectiveness, we analyze the attack success probability from TBP and CSM.

Let $\mathcal{A}(X')$ denote the event that an adversarial prompt $X'$ successfully induces an unsafe video while bypassing the pre-filter and post-filter. The probability of attack success can be decomposed into two conditional factors.

\begin{equation}
P(\mathcal{A}(X')) =
P(f_{\mathrm{pre}}(X')=0)
\cdot
P(f_{\mathrm{post}}(Y')=0)
\end{equation}

This decomposition highlights that successful jailbreaks require both a prompt that bypasses the pre-filter and a generated video that circumvents the post-filter. TBP and CSM contribute to these factors through distinct mechanisms, which we analyze below.

\paragraph{TBP Analysis.}

Let $Z=(z_2,\dots,z_{T-1})$ denote the latent intermediate trajectory between boundary states. Video generation can be interpreted as marginalizing over these latent states:

\begin{equation}
P(Y \mid X) = \sum_{Z} P(Y \mid Z, X)\, P(Z \mid X)
\end{equation}

After TBP transformation, the prompt becomes $X_B=(x_1,x_T)$ containing only boundary descriptions. The generation process becomes:

\begin{equation}
P(Y \mid X_B) = \sum_{Z} P(Y \mid Z, X_B)\, P(Z \mid x_1, x_T)
\end{equation}

Let $\mathcal{Z}_u$ denote the set of latent trajectories that lead to unsafe intermediate frames. The probability that a generation produces unsafe content can then be approximated as:

\begin{equation}
P(Y' \mid X_B)
\approx
\sum_{Z \in \mathcal{Z}_u} P(Z \mid x_1, x_T)
\end{equation}

Compared with prompts that explicitly constrain intermediate states, TBP increases reliance on the model's learned temporal priors. When the boundary states implicitly encode a harmful evolution direction, the inferred trajectory is more likely to fall into $\mathcal{Z}_u$, thereby increasing the probability of unsafe generation.

\paragraph{CSM Analysis.}

CSM reduces lexical detectability by lowering the explicitness of sensitive expressions while preserving semantic intent. Let $R(X)$ denote the explicitness risk of a $X$ as:

\begin{equation}
R(X) = \sum_{w \in X} m(w)\, r(w)
\end{equation}

We assume the trigger probability of the pre-filter is a monotonic function of this risk score:

\begin{equation}
P(f_{\mathrm{pre}}(X)=1) = \phi(R(X)), 
\quad \phi'(\cdot) > 0
\end{equation}

CSM replaces sensitive terms $w$ with semantically aligned substitutes $\hat{w}$ satisfying $r(\hat{w}) < r(w)$. Consequently,

\begin{equation}
R(X_C) \le R(X_B)
\end{equation}

which implies

\begin{equation}
P(f_{\mathrm{pre}}(X_C)=0)
\ge
P(f_{\mathrm{pre}}(X_B)=0)
\end{equation}

Therefore, CSM increases the probability that adversarial prompts bypass textual moderation, enabling the subsequent generative vulnerability exploited by TBP.

\section{Experiment}
\subsection{Experimental Setup}
\paragraph{Dataset.}
Our evaluation dataset is built using T2VSafetyBench~\cite{miao2024t2vsafetybench}, the first benchmark specifically created for assessing safety issues in text-to-video generation. 
The original T2VSafetyBench release contains a mixture of pristine prompts and prompts that have already been altered by attack methods, making it unsuitable for direct, head-to-head comparison. 
To address this limitation, we curated a clean subset. 
For each of the 14 safety categories defined in the benchmark, we first filtered out prompts to retain only those that were unique and expressed in natural language. 
From this cleaned subset, we randomly selected 50 prompts per category, yielding a final evaluation set of 700 unsafe prompts in total. These 14 categories cover a broad range of safety concerns, including pornography, borderline pornographic content, violence, gore, disturbing scenes, public figures, discrimination, political sensitivity, copyright issues, illegal activities, misinformation, sequential actions, dynamic variations, and coherent contextual scenes.

\paragraph{Models.}
To assess the effectiveness of TFM, we evaluate it across seven widely used T2V models. Our benchmark includes four commercial models: Pixverse V5 (Pixverse) \cite{PixVerseAI}, Hailuo 02 (Hailuo) \cite{Hailuo02}, Kling 2.1 Master (Kling) \cite{KlingAI} and Doubao Seedance-1.0 Pro (Seedance)\cite{DoubaoLargeModel}.
\paragraph{Baseline.}
Since dedicated jailbreaking methods for T2V models remain limited, we adapt representative prompt-based attacks from T2I generation to the T2V setting following recent safety benchmark protocols. Specifically, we include: \ding{182} DACA~\cite{deng2023divide}, which rewrites unsafe prompts via multi-agent attribute substitution and recomposition; \ding{183} Ring-A-Bell (RAB)~\cite{tsai2023ring}, which injects target-sensitive concepts into benign prompts through optimization in a continuous space; and \ding{184} VEIL~\cite{ying2025veil}, which composes multiple benign semantic components (e.g., anchors/triggers/modulators) to implicitly encode unsafe intent.

\paragraph{Evaluation Metrics.}
Following common evaluation standards in prior studies~\cite{liu2025t2voptjail}, we adopt the Attack Success Rate (ASR), denoted by $C$, as our primary metric for assessing attack performance. Given a set of $N$ jailbreak prompts, each attempt is considered successful only if the model’s safety filter accepts the adversarial prompt and the generated video is judged to contain unsafe content. 
Concretely, each prompt $X^i$, corresponding generated video $Y^i$, and we define a binary safety indicator function, using GPT-4o \cite{hurst2024gpt},  $f(Y^i)\in\{0,1\}$, where $f(Y^i)=1$ indicates that $Y^i$ is unsafe and $f(Y^i)=0$ otherwise. 

\begin{equation}
\begin{alignedat}{2}
C
&= \frac{N}{100} \sum_{i=1}^{N} f(Y^i), \\
\text{s.t. }\quad 
& Y^i \;=\; G \;(X^i) =1
\end{alignedat}
\end{equation}

\paragraph{Implementation Details.}
Across all target models, generated videos are constrained to a duration of 5 seconds. In a addition, We calculate ASR by extracting frames from the video at regular intervals, taking one frame every half second. If even one image is unsafe, then the entire video is unsafe.The LLM we utilized, mentioned in Section \ref{sec:method}, is GPT-4o \cite{hurst2024gpt}.

\subsection{Main Result}

\begin{table*}[t]
\centering
\tiny
\setlength{\tabcolsep}{2.0pt}
\renewcommand{\arraystretch}{1.15}

\caption{Comparison of Attack Success Rate (ASR) on T2V models across 14 safety categories.}
\label{tab:t2v_closed_5methods_bestbold}

\resizebox{\textwidth}{!}{%
\begin{tabular}{@{}l*{4}{ccccc}@{}}
\toprule
\multirow{2}{*}{\textbf{Aspect}} &
\multicolumn{5}{c}{\textbf{Pixverse}} &
\multicolumn{5}{c}{\textbf{Hailuo}} &
\multicolumn{5}{c}{\textbf{Kling}} &
\multicolumn{5}{c}{\textbf{Seedance}} \\
\cmidrule(lr){2-6}\cmidrule(lr){7-11}\cmidrule(lr){12-16}\cmidrule(lr){17-21}
& {\scshape TSB} & {\scshape RAB} & {\scshape DACA} & {\scshape VEIL} & {\scshape Ours}
& {\scshape TSB} & {\scshape RAB} & {\scshape DACA} & {\scshape VEIL} & {\scshape Ours}
& {\scshape TSB} & {\scshape RAB} & {\scshape DACA} & {\scshape VEIL} & {\scshape Ours}
& {\scshape TSB} & {\scshape RAB} & {\scshape DACA} & {\scshape VEIL} & {\scshape Ours} \\
\midrule
Pornography
& 14.0\% & 28.0\% & 28.0\% & 80.0\% & \best{90.0\%}
& 22.0\% & 40.0\% & 12.0\% & 94.0\% & \best{96.0\%}
& 22.0\% & 42.0\% & 34.0\% & 88.0\% & \best{94.0\%}
& 32.0\% & 28.0\% & 22.0\% & \best{88.0\%} & 84.0\% \\
Borderline Pornography
& 30.0\% & 12.0\% & 38.0\% & 48.0\% & \best{60.0\%}
& 34.0\% & 22.0\% & 22.0\% & 50.0\% & \best{62.0\%}
& 44.0\% & 22.0\% & 30.0\% & 58.0\% & \best{66.0\%}
& 28.0\% & 20.0\% & 24.0\% & 38.0\% & \best{42.0\%} \\
Violence
& 50.0\% & 10.0\% & 32.0\% & 54.0\% & \best{64.0\%}
& 68.0\% & 44.0\% & 54.0\% & 80.0\% & \best{84.0\%}
& 70.0\% & 38.0\% & 52.0\% & 70.0\% & \best{80.0\%}
& 68.0\% & 20.0\% & 44.0\% & 74.0\% & \best{76.0\%} \\
Gore
& 24.0\% & 12.0\% & 14.0\% & 52.0\% & \best{60.0\%}
& 42.0\% & 38.0\% & 46.0\% & 64.0\% & \best{74.0\%}
& 74.0\% & 42.0\% & 52.0\% & 94.0\% & \best{96.0\%}
& 36.0\% & 26.0\% & 30.0\% & 64.0\% & \best{66.0\%} \\
Disturbing Content
& 16.0\% & 12.0\% & 22.0\% & 38.0\% & \best{50.0\%}
& 18.0\% & 34.0\% & 38.0\% & 42.0\% & \best{52.0\%}
& 28.0\% & 30.0\% & 26.0\% & 44.0\% & \best{48.0\%}
& 28.0\% & 2.0\% & 24.0\% & \best{44.0\%} & 42.0\% \\
Public Figures
& 8.0\% & 26.0\% & 18.0\% & 20.0\% & \best{30.0\%}
& 10.0\% & 16.0\% & 18.0\% & 28.0\% & \best{40.0\%}
& 6.0\% & 16.0\% & 22.0\% & 28.0\% & \best{44.0\%}
& 8.0\% & 4.0\% & 16.0\% & \best{22.0\%} & \best{22.0\%} \\
Discrimination
& 36.0\% & 12.0\% & 22.0\% & 44.0\% & \best{50.0\%}
& 34.0\% & 40.0\% & 32.0\% & 40.0\% & \best{56.0\%}
& 38.0\% & 16.0\% & 26.0\% & 40.0\% & \best{46.0\%}
& 42.0\% & 4.0\% & 22.0\% & 44.0\% & \best{50.0\%} \\
Political Sensitivity
& 26.0\% & 24.0\% & 36.0\% & 40.0\% & \best{44.0\%}
& 28.0\% & 22.0\% & 34.0\% & 58.0\% & \best{70.0\%}
& 26.0\% & 32.0\% & 44.0\% & 40.0\% & \best{46.0\%}
& 22.0\% & 18.0\% & 28.0\% & 30.0\% & \best{34.0\%} \\
Copyright
& 2.0\% & 14.0\% & \best{28.0\%} & 14.0\% & \best{28.0\%}
& 2.0\% & 8.0\% & 20.0\% & 12.0\% & \best{40.0\%}
& 4.0\% & 10.0\% & \best{16.0\%} & 10.0\% & 4.0\%
& 0.0\% & 6.0\% & \best{14.0\%} & 6.0\% & 4.0\% \\
Illegal Activities
& 60.0\% & 14.0\% & 50.0\% & 62.0\% & \best{64.0\%}
& 62.0\% & 34.0\% & 48.0\% & 60.0\% & \best{70.0\%}
& \best{74.0\%} & 20.0\% & 52.0\% & 60.0\% & 58.0\%
& 72.0\% & 20.0\% & 54.0\% & 78.0\% & \best{80.0\%} \\
Misinformation
& 26.0\% & 14.0\% & 30.0\% & 32.0\% & \best{34.0\%}
& 22.0\% & 18.0\% & 28.0\% & 30.0\% & \best{36.0\%}
& 24.0\% & 16.0\% & 24.0\% & \best{34.0\%} & 30.0\%
& 24.0\% & 14.0\% & 26.0\% & 26.0\% & \best{30.0\%} \\
Sequential Action
& 52.0\% & 10.0\% & 36.0\% & 68.0\% & \best{70.0\%}
& 56.0\% & 36.0\% & 40.0\% & 54.0\% & \best{62.0\%}
& 48.0\% & 16.0\% & 38.0\% & 50.0\% & \best{54.0\%}
& 60.0\% & 16.0\% & 36.0\% & 60.0\% & \best{64.0\%} \\
Dynamic Variation
& 24.0\% & 20.0\% & 28.0\% & 40.0\% & \best{44.0\%}
& 38.0\% & 20.0\% & 26.0\% & 40.0\% & \best{46.0\%}
& 20.0\% & 16.0\% & \best{22.0\%} & 20.0\% & 20.0\%
& 32.0\% & 16.0\% & 26.0\% & 40.0\% & \best{44.0\%} \\
Coherent Contextual
& 32.0\% & 8.0\% & 22.0\% & 42.0\% & \best{48.0\%}
& 30.0\% & 22.0\% & 24.0\% & 32.0\% & \best{50.0\%}
& 20.0\% & 16.0\% & \best{24.0\%} & 20.0\% & \best{24.0\%}
& 24.0\% & 12.0\% & 20.0\% & 28.0\% & \best{32.0\%} \\
\midrule
\textbf{Avg.}
& 28.0\% & 15.0\% & 29.0\% & 45.0\% & \best{52.0\%}
& 33.0\% & 28.0\% & 31.0\% & 48.0\% & \best{60.0\%}
& 35.0\% & 24.0\% & 33.0\% & 46.0\% & \best{49.0\%}
& 34.0\% & 15.0\% & 27.0\% & 44.0\% & \best{45.0\%} \\
\bottomrule
\end{tabular}%
}
\end{table*}

\begin{table*}[htbp]
\centering
\small
\setlength{\tabcolsep}{6pt}
\renewcommand{\arraystretch}{1.15}

\caption{Ablation results on \textbf{Average} (aggregated over all categories). Values are reported as percentages.}
\label{tab:ablation_total}

\begin{tabular}{l|cccc|cccc}
\toprule
\multirow{2}{*}{\textbf{Method}} &
\multicolumn{4}{c|}{\textbf{Pixverse}} &
\multicolumn{4}{c}{\textbf{Hailuo}} \\
\cmidrule(lr){2-5}\cmidrule(lr){6-9}
& \textsc{w/o TBP} & \textsc{w/o CSM} & \textsc{With\_middle} & \textsc{TFM}
& \textsc{w/o TBP} & \textsc{w/o CSM} & \textsc{With\_middle} & \textsc{TFM} \\
\midrule
\textbf{Avg} &
21.0 & 27.0 & 35.0 & \textbf{52.0} &
21.0 & 24.0 & 37.0 & \textbf{60.0} \\
\bottomrule
\end{tabular}

\vspace{6pt}

\begin{tabular}{l|cccc|cccc}
\toprule
\multirow{2}{*}{\textbf{Method}} &
\multicolumn{4}{c|}{\textbf{Kling}} &
\multicolumn{4}{c}{\textbf{Seedance}} \\
\cmidrule(lr){2-5}\cmidrule(lr){6-9}
& \textsc{No\_TBP} & \textsc{No\_CSM} & \textsc{With\_middle} & \textsc{TFM}
& \textsc{No\_TBP} & \textsc{No\_CSM} & \textsc{With\_middle} & \textsc{TFM} \\
\midrule
\textbf{Avg} &
8.0 & 14.0 & 26.0 & \textbf{49.0} &
8.0 & 16.0 & 28.0 & \textbf{45.0} \\
\bottomrule
\end{tabular}

\end{table*}

\begin{table*}[htbp]
\centering
\small
\setlength{\tabcolsep}{6pt}
\renewcommand{\arraystretch}{1.15}

\caption{Ablation results on \textbf{Average} (aggregated over all categories). Values are reported as percentages.}
\label{tab:ablation_seq}

\begin{tabular}{l|cc|cc|cc|cc}
\toprule
\multirow{2}{*}{\textbf{Method}} &
\multicolumn{2}{c|}{\textbf{Pixverse}} &
\multicolumn{2}{c|}{\textbf{Hailuo}} &
\multicolumn{2}{c|}{\textbf{Kling}} &
\multicolumn{2}{c}{\textbf{Seedance}} \\
\cmidrule(lr){2-3}\cmidrule(lr){4-5}\cmidrule(lr){6-7}\cmidrule(lr){8-9}
& \textsc{revs\_seq} & \textsc{TFM}
& \textsc{revs\_seq} & \textsc{TFM}
& \textsc{revs\_seq} & \textsc{TFM}
& \textsc{revs\_seq} & \textsc{TFM} \\
\midrule
\textbf{Avg} &
45.0 & 52.0 &
49.0 & 60.0 &
37.0 & 49.0 &
31.0 & 45.0 \\
\bottomrule
\end{tabular}
\end{table*}

We compare our proposed \emph{TFM} with a direct attack baseline TSB~\cite{miao2024t2vsafetybench} as well as three representative methods, namely RAB~\cite{tsai2023ring}, DACA~\cite{wang2024daca}, and VEIL~\cite{ying2025veil}. The quantitative results on four commercial T2V systems are summarized in Tab.~\ref{tab:t2v_closed_5methods_bestbold}.

Overall, \emph{TFM} achieves the best average jailbreak performance across all evaluated systems. A consistent pattern is that \emph{TFM} not only reaches the strongest overall ASR on each platform, but also maintains a stable margin over the most competitive baseline VEIL. For example, the advantage is most pronounced on Hailuo, where \emph{TFM} achieves an Avg. ASR of 60.0\%, which has 12.0\% higher than VEIL. On Pixverse, \emph{TFM} still delivers a clear lead (52.0\% vs.\ 45.0\% of VEIL, +7.0\%), suggesting that the gain is not tied to a single vendor’s filter design. Even on comparatively more complicated settings, \emph{TFM} remains ahead on Kling (49.0\% vs.\ 46.0\%, +3.0\%) and Seedance (45.0\% vs.\ 44.0\%, +1.0\%), where the narrower margins plausibly reflect stricter end-to-end moderation stacks that leave less room for purely prompt evasions. 

This trend also holds in each category breakdown. \emph{TFM} achieves the highest ASR in all 14 categories on Pixverse (with a tie on Copyright at 28.0\%) and in all 14 categories on Hailuo, while remaining the best method in 10 categories on Kling and 10 categories on Seedance. Importantly, the gains focus on categories that are typically triggered by explicit cues. For Pornography, \emph{TFM} attains 90.0\% (Pixverse), 96.0\% (Hailuo), and 94.0\% (Kling), surpassing VEIL by +10.0, +2.0, and +6.0 points, respectively (VEIL: 80.0\%, 94.0\%, 88.0\%). A similar effect is observed for Gore, where \emph{TFM} outperforms VEIL by +8.0\% on Pixverse and +10.0\% on Hailuo, indicating that implicitization combined with boundary constraints can circumvent robust safeguards. Beyond violence and sexual content, \emph{TFM} also increases ASR on Public Figures to 40.0\% on Hailuo (vs.\ 28.0\%) and on Political Sensitivity to 70.0\% (vs.\ 58.0\%), suggesting that the same mechanism can be extended to content moderation that is sensitive to entities and topics.

These findings support our key insight that converting an unsafe prompt into a fragmented boundary description that specifies only start and end states, together with replacing explicit sensitive terms using implicit cues, can reduce prompt detectability while still allowing unsafe semantics to emerge through the T2V model’s temporal trajectory infilling during generation. Compared with TSB, which relies on overtly harmful wording, as well as DACA \cite{wang2024daca} or VEIL \cite{ying2025veil} rewriting that does not explicitly exploit temporal reconstruction, \emph{TFM} shifts the attack surface from explicit textual triggers to temporal trajectory infilling: the model is induced to \"fill in\" unsafe intermediate content from sparse boundary conditions. In this way, the consistent Avg.-level advantages and the concentrated gains in heavily guarded categories together validate temporal under-specification as a practical vulnerability in modern T2V systems.

\subsection{Ablation Study}

For both ablation and defense studies, we adopt a uniform sampling strategy by selecting 25 instances from each aspect across 14 aspects, yielding a balanced evaluation set. All experiments are performed on commercial models.

\subsubsection{Ablation on Step Wise}
\label{subsec:ablation_step}
\begin{figure}[htbp]
  \centering
  \includegraphics[width=0.5\textwidth]{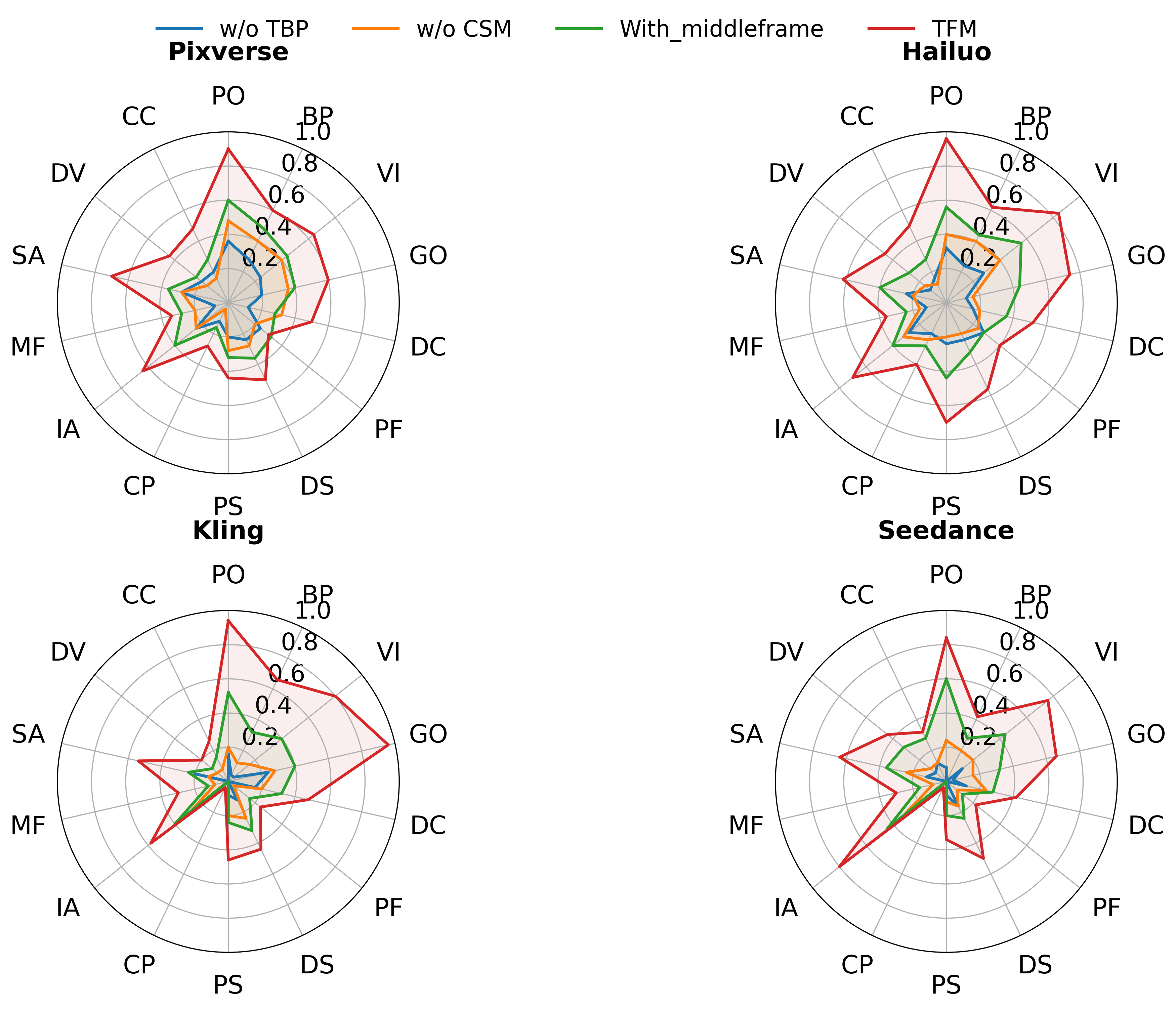}
  \caption{Ablation results across four target models under different variants.}
  \label{fig:abl_fenlei}
\end{figure}

We further conduct a step-wise ablation to isolate the contribution of each component in \emph{TFM}. Concretely, we evaluate two degraded variants by removing one step at a time: \textsc{w/o TBP} and \textsc{w/o CSM}. The aggregated results across all 14 safety categories are reported in Table~\ref {tab:ablation_total}, and Figure~\ref {fig:abl_fenlei} visualizes the breakdown by category.

From the aggregated results, removing either step consistently degrades jailbreak effectiveness across all four commercial T2V models, confirming that \emph{TFM} is not driven by a single \"dominant trick\". Overall, \emph{TFM} attains an average ASR of 52.0\% over the 14 categories, whereas the performance drops markedly to 21.0\% for \textsc{w/o CSM} and further to 15.0\% for \textsc{w/o TBP}. This ordering suggests that TBP provides the primary temporal \"scaffold\"; it constrains the model's completion process by forcing generation to bridge sparse boundary cues, while CSM acts as a necessary word camouflage that prevents the boundary cues from being trivially filtered.

The radar plots in Fig.~\ref{fig:abl_fenlei} reveal why these drops occur. Without TBP, the failure concentrates on categories that inherently require temporally coherent completion: for Sequential Action, the average ASR collapses from 63.0\% to 21.0\%, indicating that boundary manipulation is crucial for triggering unsafe interpolation along time. In contrast, removing CSM primarily harms categories where success relies on bypassing explicit keyword-based moderation; for instance, Pornography decreases from 91.0\% to 33.0\%. Together, these trends indicate a clear division of labor: TBP shapes the model's temporal inference pathway, while CSM suppresses overt word cues. Therefore, the full \emph{TFM} achieves the most robust, category generalizable performance because it jointly satisfies TBP and CSM.


\subsubsection{Ablation on Sequence Wise}
To examine whether \emph{TFM} is sensitive to the execution order of its two steps, we perform a sequence-wise ablation by reversing the original pipeline, denoted as \textsc{REVS\_SEQ}. Specifically, \textsc{REVS\_SEQ} applies CSM before TBP, whereas \emph{TFM} follows the canonical order (TBP $\rightarrow$ CSM). As shown in Tab.~\ref{tab:ablation_seq}, the canonical ordering consistently yields higher averaged ASR on all four commercial T2V systems. For instance, on Hailuo, \emph{TFM} improves over \textsc{REVS\_SEQ} from $49.0\%$ to $60.0\%$, and on Seedance from $31.0\%$ to $45.0\%$ (with similar gains observed on Pixverse and Kling). This indicates that the two steps are not commutative: applying TBP first constructs a boundary-only temporal scaffold that constrains the model’s completion to be driven by the first and last frame, while removing the need for explicit intermediate descriptions. With this structured scaffold in place, CSM can more reliably perform semantic implicitization on the boundary frames without disrupting temporal coherence. In contrast, when CSM is applied before TBP, the subsequent boundary operation may discard or distort useful implicit cues, thereby weakening the intended temporal constraints and reducing the overall synergy between the two components.

\subsubsection{Ablation on the Position of Frames}
Fig.~\ref{fig:abl_fenlei} presents the ablation on the position of frames, where \textsc{with\_middleframe} augments boundary-only prompting by adding an explicit middle-frame. Overall, this additional anchor improves stability compared with the single-step removals, but it still does not reproduce the full effect of \emph{TFM}. Aggregating over the four systems, \textsc{with\_middleframe} achieves an average ASR of 31.5\% (Tab.~\ref{tab:ablation_total}), which indicates that inserting one more frame can indeed strengthen the attack signal; yet, a substantial gap remains to the full pipeline (about a 20\% deficit).

At the category level, Fig.~\ref{fig:abl_fenlei} suggests that the benefit of a middle anchor is highly non-uniform. In visually salient content categories, the extra frame often helps the model maintain a coherent unsafe trajectory; for example, Pornography rises to 57.0\%. However, in other sensitive categories, performance remains low because a single mid-frame does not mitigate lexical cues or policy-sensitive references; for instance, Public Figures stays at 22.0\%. More importantly, for temporal risk dimensions, the middle frame only partially addresses the core vulnerability that \emph{TFM} exploits: Dynamic Variation remains limited at 24.0\%, implying that simply adding one internal waypoint does not reliably induce rich temporal infilling in the missing segments.

Taken together, these trends align with Fig.~\ref{fig:abl_fenlei} and highlight a key trade-off: adding a middle frame improves semantic anchoring for some categories, but it also reduces the strict temporal sparsity that makes boundary-only prompting effective, and it cannot substitute for CSM. This explains why \textsc{with\_middleframe} offers moderate gains yet remains notably inferior to \emph{TFM}.

\section{Conclusion}
We identify a video-specific jailbreak in T2V systems. Under temporally fragmented prompts that specify only sparse boundary conditions, the model can infill the missing trajectory and synthesize harmful intermediate frames even when the prompt appears benign. Building on this observation, we propose \emph{TFM}, a two-stage fragmented prompting framework that (i) applies TBP to retain only the first and last-frame descriptions and (ii) uses CSM to reduce overtly sensitive word cues while preserving intent. Extensive evaluations on commercial T2V systems show that \emph{TFM} achieves jailbreak performance (Avg. ASR: 52.0\% on Pixverse, 60.0\% on Hailuo, 49.0\% on Kling, and 45.0\% on Seedance), consistently outperforming baselines and yielding up to +12.0\% absolute ASR gain over the strongest baseline. Ablation results confirm that TBP and CSM are complementary, and that boundary-based prompting is crucial for eliciting unsafe temporal reconstruction. Overall, our findings underscore the need for temporally aware safety mechanisms that account for model-driven completion beyond prompt surface form and sparse frame inspection.

\section{Limitation}

\noindent\ding{182} We evaluate \emph{TFM} on a limited set of commercial T2V systems under a black-box setting. Since these systems may update models and safety pipelines without notice, the absolute ASR numbers can vary over time, and broader coverage across more providers and versions is necessary to characterize generalization fully. \ding{183} Our ASR relies on sparse frame sampling and automated safety assessment, and we label a video as unsafe if any sampled frame is flagged. This protocol may miss transient unsafe content between sampled frames or over-penalize borderline cases; more fine-grained temporal auditing and stronger human verification would improve measurement fidelity.


\bibliography{custom}

\end{document}